\documentclass[aps,pra,showpacs,twocolumn,superscriptaddress]{revtex4}
\usepackage{amsmath}    
\usepackage{amsfonts}
\usepackage{graphicx}   
\usepackage{dcolumn}
\usepackage{bm}
\usepackage{color}      
\usepackage{subfigure}    
\newcommand{\ket}[1]{\ensuremath{|{#1}\rangle}}
\newcommand{\pstate}[1]{\ensuremath{|{#1}\rangle\langle{#1} |}}
\newcommand{\bra}[1]{\ensuremath{\langle{#1} |}}

\newcommand{\beq}{\begin{equation}}
\newcommand{\eeq}{  \end{equation}}
\newcommand{\bea}{\begin{eqnarray}}
\newcommand{\eea}{\end{eqnarray}}
\newcommand{\beqn}{\begin{eqnarray}}
\newcommand{\eeqn}{\end{eqnarray}}
\newcommand{\bit}{\begin{itemize}}
\newcommand{\eit}{  \end{itemize}}

\begin{document}


\title{Decoherence, entanglement decay, and equilibration produced by chaotic environments.}

\author{Gabriela Barreto Lemos}
\email{gabibl@if.ufrj.br}
\affiliation{Instituto de F\'{\i}sica, Universidade Federal do Rio
de Janeiro, Caixa Postal 68528, Rio de Janeiro, RJ 21941-972,
Brazil}

\author{Fabricio Toscano}
\affiliation{Instituto de F\'{\i}sica, Universidade Federal do Rio de
Janeiro, Caixa Postal 68528, Rio de Janeiro, RJ 21941-972, Brazil}

\begin{abstract}
We investigate decoherence in quantum systems coupled via dephasing-type 
interactions to an arbitrary environment with 
chaotic underlying classical dynamics.
The coherences of the reduced state of the central system written in the preferential energy eigenbasis are quantum Loschmidt echoes, which in the strong coupling regime are characterized at long times scales by fluctuations around a constant mean value. We show 
that due to the chaotic dynamics of the environment, the mean value and the
width of the Loschmidt echo fluctuations are inversely proportional to the quantity we define as the effective Hilbert space dimension of the environment,  
which in general is smaller than the dimension of the entire available Hilbert space. 
Nevertheless, in the semiclassical regime this effective Hilbert space dimension is in general large, in which case even a chaotic environment with few degrees of freedom produces decoherence without revivals.
Moreover we show that in this regime the environment  leads the central system to equilibrate to the time average of its reduced density matrix, which corresponds to a diagonal state in the preferential energy eigenbasis.
For the case of two uncoupled, initially entangled central systems that interact with identical local quantum environments with chaotic underlying classical dynamics, 
we show that  in the semiclassical limit the equilibration state 
is arbitrarily close to a separable state.  
We confirm our results with numerical simulations in which the environment is modeled by the quantum kicked rotor in the chaotic regime.
\end{abstract}

\pacs{}

\maketitle

\section{Introduction}
\label{sec:introduction}
In open quantum systems the interaction between a system and its environment may result in the well known phenomenon of decoherence~\cite{zurek2003,petruccione}. While the central system becomes ever more entangled with the environment, quantum information initially present in the reduced state of the system may be lost to the environment. 
The most transparent example of how the irreversible loss of quantum information to the environment leads to decoherence is in the case of dephasing-type system-environment interactions (or so-called \textit{measurement-type} interactions\cite{petruccione}), {\it i.e.,} when the interaction Hamiltonian commutes with the free system Hamiltonian.
In this case, by tracing over environmental degrees of freedom, one may observe the irreversible decay of the quantum coherences of the central system's reduced density
matrix written in the preferential basis of the free Hamiltonian eigenstates, while the populations are conserved. 

The traditional approach to this problem points to the need for an infinite number of environmental degrees of freedom in order for the decoherence process to occur~\cite{petruccione}. The Caldeira-Leggett model \cite{caldeira-leggett} is the most renowned environment model of this
type. However, recently there has been an increasing interest in understanding if and how environments with few degrees of freedom 
can produce decoherence in a quantum system of interest 
\cite{kubotani,doron-cohen,kohout,lee, aguiar2, rossini,saraceno,Pineda2006, Fonseca2008,casati2008, bandyop, buric,jacquod,oliveira2009,lemos2010}
and if decoherence can be produced by few internal degrees of freedom~\cite{sokolov2007,kim}.
One motivation for this is the relevance of decoherence to quantum computation and quantum information tasks, where often the interaction of the system of interest with a global environment (usually composed of many degrees of freedom) is well screened and therefore the interaction with a ``near"  
environment composed of few degrees of freedom involved in the control of quantum operations may be the  most relevant \cite{gorin2008, Benenti2009}. 
The other motivation is the study of the emergence of classicality in quantum systems, where decoherence plays a central role.

The study of environments with few degrees of freedom is in general related to the role of 
the chaotic dynamics of their classical counterpart \cite{kubotani,doron-cohen,kohout,lee, aguiar2, rossini,saraceno,Pineda2006, Fonseca2008,casati2008, bandyop, buric,jacquod,lemos2010}\footnote{Due to the well-established connection of random matrix theory and quantum chaos 
some studies use a random matrix model of the environment  
\cite{gorin2008}.}.
The connection between {\it environment-induced decoherence} and the so called {\it fidelity} decay (or quantum Loschmidt-echo decay
\footnote{The Loschmidt echo is related to the procedure of propagating a system forward in time with some Hamiltonian and then back with a perturbed one (see \cite
{gorin2006} and references therein). 
The overlap between the initial pure state and the final one, after forward and backward evolution,  is
called  \textit{fidelity}  \cite{peres1984}, and measures the sensitivity of the system to perturbations. A similar quantity for mixed states (the {\it allegiance}) was introduced in \cite{sokolov2007}.})  brought great 
insight into the study of decoherence by chaotic environments \cite{butterfly,zurek2003,Cucchietti2003,gorin2004,sokolov2007,jacquod,prosen2003}. 
The sensitivity of chaotic environments to the
perturbations produced by the interaction with a central system was conjectured in \cite{butterfly}
to be related to their ability to rapidly produce decoherence. 
A precise connection  between quantum Loschmidt echo and  decoherence was first made in 
\cite{gorin2004} in the case of dephasing-type system-environment interactions, where the 
off-diagonal elements of the system's reduced matrix written in the preferential basis are 
different Loschmidt echoes in the environmental degrees of freedom.
This connection does not neglect the free evolution of the central system. 
A similar approach was developed in \cite{casati2008}, but for a generic small coupling to the environment.
In that article the authors show, using perturbative approximations, that the 
off-diagonal elements of the reduced matrix of the system
in its energy eigenstates basis are proportional  to {\it fidelities}.

It is common to associate the decoherence time with the characteristic 
short-time  scale for the decay
of the {\it fidelity} \cite{casati2008, butterfly}. However, for long times the {\it fidelity} in general
can suffer fluctuations that can be large and may lead to important revivals of the coherences.    
Indeed, if the quantum environment has chaotic underlying classical 
dynamics, at long-time scales the coherences of the central system's reduced density matrix, written in its preferential eigenbasis, fluctuate around a constant mean value.  For a generic central system strongly coupled by dephasing-type interactions to an arbitrary chaotic environment (which may have few degrees of freedom), we show that the time-average and the width of these fluctuations are inversely proportional to the \textit{effective dimension} of the environment's Hilbert space, which can always be defined (even in the case in which the total available Hilbert space
of the environment has infinite dimension) .
Thus, for chaotic environments, the decoherence occurs in the semiclassical limit, 
{\it i.e.,} $\hbar_{\rm eff}=\hbar/S\rightarrow 0$  ($S$ is a typical action of the environment), 
where typically the effective Hilbert space dimension of the environment is large. 

Our results have a direct application to the problem of equilibration, a key process in the understanding of  thermal equilibration in quantum systems \cite{Popescu2006,Linden2009,zanardi2010}.
A central system equilibrates if its initial state evolves toward some particular state, in general mixed, 
and remains in that state, or close to it, for all times. 
Here we show that if the environment Hamiltonian has underlying classical chaotic dynamics, 
and for a dephasing system-environment interaction, the central system equilibrates to a totally decohered
mixed state in the semiclassical regime, independent of the initial state of the system 
and for generic initial states of the environment (initially decoupled from the system of interest). 

 The study of entanglement decay due to the action of environments 
is a central problem in quantum computation and quantum-information processing.
In this regards it is important to know how chaotic environments with few degrees of freedom can produce entanglement decay in the system of interest~\cite{ rossini, Pineda2006, bandyop, buric, lemos2010}.
Here we show that  an initially entangled state of  two non-interacting central systems coupled to equivalent local quantum environments with chaotic classical dynamics  equilibrates to a state that  is arbitrarily close to a separate state in the semiclassical limit.

The paper is organized as follows: In Sec.\ref{sectionII} we present the echo dynamics approach to decoherence. We then obtain the long-time behavior of decoherence functions for chaotic environments in Sec.\ref{sectionIII} and go on to discuss decoherence and equilibration of a one party central system in Sec.\ref{sectionIV}. Disentanglement and equilibration of bipartite systems produced by chaotic 
environments are discussed in Sec.\ref{sectionV} and in Sec.\ref{sectionVI} we describe our numerical simulations and analyze their results. We finish with concluding remarks in  Sec.\ref{sectionVII}. 

\section{Echo dynamics approach to decoherence}
\label{sectionII}

We consider an arbitrary system interacting with a generic dephasing environment:
\beq
\label{general-model-1}
\hat H= \hat H_c\otimes\hat\openone_e  +g\hat S \otimes  \hat  V + \hat\openone_c\otimes\hat H_e,
\eeq 
where  $\hat H_c$ and $\hat H_e$  are, respectively, the Hamiltonians of the central system 
and the environment. The dephasing interaction is 
$\hat H_I=g \hat S \otimes \hat  V $, where $\hat S$ acts on the system degrees of freedom and
$[ \hat H_c, \hat S]=0$. The operator $\hat V$ acts on environmental degrees of freedom, and $g$ is the coupling strength. 
The global initial state is a product state:
\beq
\label{uncoupled-initial-state}
\hat \rho_{ce}(0)=\hat \rho(0) \otimes \hat \omega(0) =
\sum_{n,m} A_{nm} \ket n\bra m \otimes \hat \omega(0)\;\;,
\eeq
where $\hat\omega(0)$ is the initial reduced state of the environment, and we have expanded the initial state of the central system $\hat\rho(0)$ in the preferential basis of common 
eigenstates of $ \hat H_c$ and $\hat S$, which we assume to have discrete and non-degenerate spectra:
\bea
\hat H_c \ket n &=& \varepsilon_n \ket n, \nonumber \\
\hat S \ket n &=& s_n \ket n.
\eea 

The evolved reduced 
density matrix of the central system is obtained by tracing over the environmental degrees of freedom
\beq
\label{rdm}
\hat \rho(t)=
\sum_{n,m} A_{nm} e^{-i(\varepsilon_n-\varepsilon_m)t/\hbar}{\rm Tr}_e\left[\hat U_m^{\dagger}\hat U_n \hat \omega(0) \right]\ket n\bra m,
\eeq
where $\hat U_{I+e,n(m)}\equiv\hat U_{n(m)}$ are conditional effective 
evolution operators associated with the Hamiltonians 
\begin{equation}
\label{unperturbed-Hamil}
\hat H_{n(m)}=\hat H_e + gs_{n(m)} \hat V, 
\end{equation}
which act exclusively on environmental degrees 
of freedom.
Hence, the central system's diagonal matrix elements in the preferred basis $\{\ket n\}$ 
are constant, while the time evolution of off-diagonal matrix elements is controlled
by the decoherence functions~\cite{petruccione} 
\beq
\label{dec}
F_{nm}(t)=\left\vert{\rm Tr}_e\left[\hat U_m(-t)\hat U_n(t) \hat \omega(0) \right]\right\vert^2,n\neq m.
\eeq

Following essentially the approach in ~\cite{gorin2004} (see also ~\cite{butterfly, casati2008,jacquod}), we can associate the decay of these 
decoherence functions with the decay of quantum Loschmidt echoes in the Hilbert space of the environment.
Indeed, the amplitudes of the decoherence functions (\ref{dec}) are the so-called \textit{allegiance} amplitudes \cite{sokolov2007},
\beq
\label{fidelity1}
f_{{nm}}(t)={\rm Tr}_e\left[ \hat M_{{nm}}(t) \hat \omega(0) \right]\;\;,
\eeq 
where we have introduced the echo operator that acts in the Hilbert space of the
environment,
\begin{eqnarray}
\label{echo-operator}
\hat M_{nm}(t)&=&\hat U_m(-t)\hat U_n(t).
\end{eqnarray} 

In this echo dynamics, $\hat H_m$ \eqref{unperturbed-Hamil} plays the role of the unperturbed Hamiltonian, and $\hat H_n$ \eqref{unperturbed-Hamil}, rewritten as 
$H_n=\hat H_m+\epsilon_{nm}\hat V$, is the perturbed Hamiltonian with perturbation amplitude given by $\epsilon_{nm}\equiv g(s_n-s_m)$.

For pure initial states $\hat \omega(0)$ the {allegiance} amplitudes in Eq.(\ref{fidelity1}) reduce to 
the so-called fidelity amplitudes introduced by Peres \cite{peres1984}. 
It is important to note that the decay of an off-diagonal element in a given column of the system's reduced density matrix is determined by an echo operator
composed of an unperturbed evolution operator $\hat U_m$, which is common to all the elements in that column and a perturbed operator $\hat U_n$. The
 perturbation strength $\epsilon_{nm}$ increases as we move away from the diagonal. 
Thus, all echo dynamics associated to a given column $m$ are in the strong perturbation regime
if and only if  $\epsilon_{m+1,m}\hat V$ represents a strong 
perturbation. 

In general, the time evolution of the  $F_{nm}(t)\equiv |f_{{nm}}(t)|^2$  depends
on the specific system-environment coupling, on the environment Hamiltonian $\hat H_e$, and 
on the properties of the initial state $\hat \omega(0)$. 
Complete decoherence is said to have occurred if after a typical 
time scale 
, called the decoherence time, 
the system's reduced density matrix becomes diagonal in the preferred basis. 
Nevertheless, we can also say that the central system suffers decoherence or partial decoherence 
if in the long-time regime, $F_{nm}(t)$ decreases to very 
small values and remains small for  
times much longer than the decoherence time (the central system never regains its coherences). 
This means that in order to produce true decoherence on another system, 
 the environment echo dynamics must decrease significantly in the long-time regime and 
must not present revivals thereafter.

A very simple example of a rapid initial decrease of $F_{nm}(t)$ which presents revivals in the echo dynamics consists of an arbitrary central system coupled
in the form  
$\hat H_I= \hat H_c \otimes \sum_{j=1}^{N} (g_j\hat b_j^{\dagger}+g^*_j \hat b_j)$ 
to a non-chaotic environment composed of $N$ bosonic modes described by a set of harmonic oscillators, 
$\hat H_e= \sum_{j=1}^{N}\hbar \omega_j \hat b_j^{\dagger}\hat b_j$.
When the initial state of the environment is 
$\hat \omega(0)=\pstate 0 \otimes \ldots \otimes \pstate 0$ (where $\ket 0$ is the vacuum state),
and for a constant spectral density of environment modes with a finite cut-off,
 it is straightforward to show that the system coherence amplitudes will oscillate in time. So, at finite times the central system regains its lost coherence.
In order for decoherence to occur, without revivals, one must perform the continuous 
limit of environment modes. 
In this case the decoherence functions are given by  exponentially decreasing functions with no revivals.
Therefore, in this simple example, the environment can produce true decoherence on the condition that it has a very large number of degrees of freedom.

In the case of an environment with few degrees of freedom, we will show that a 
very similar situation arises when 
the echo dynamics in the environment is associated with a Hamiltonian which has  chaotic underlying classical dynamics. 
 In this case decoherence (with no revivals) is produced when the effective dimension of the environment's Hilbert space is very large, a condition that generically can be satisfied in the semi-classical regime (small effective Planck constant), independent of the system-environment initial state. 

\section{long-time behavior of decoherence functions $F_{nm}(t)$
for chaotic environments}
\label{sectionIII}

In this section we 
determine the long-time behavior of 
the decoherence functions $F_{nm}(t)$ in the case of chaotic environments.
When the classical dynamics  associated with the free environment Hamiltonian 
$\hat H_e$ is fully chaotic, 
one can generally expect that the classical dynamics associated with the effective Hamiltonian
$\hat H_m$ [Eq.(\ref{unperturbed-Hamil})] is also fully chaotic. 
In this case, the behavior in time of the decoherence functions $F_{nm}(t)$, 
determined by the echo operator in Eq.(\ref{echo-operator}), consists 
essentially in an exponential decay
with different decay rates depending on the perturbation regime \cite{gorin2006, sokolov2007}.
However, when the Hilbert space of the environment is finite, the 
decoherence functions will not decay to zero even at long-time scales. 
Indeed, 
the discrete spectrum of the evolution operators in the echo operator
in Eq.(\ref{echo-operator}) causes $F_{nm}$ to fluctuate around the 
time-average value $\langle F_{nm}\rangle$~\cite{gorin2006,bookbenenti}, where  
\beq
\label{time-averaged}
\langle\bullet\rangle= \lim_{t \rightarrow \infty}\frac{1}{t}\int^t_0 \bullet \; dt^{'}.
\eeq 
Nevertheless, even if the Hilbert space of the environment is infinite, the
effective Hilbert space covered by the evolved state of the environment
is always limited, in which case $\langle F_{nm}\rangle$ also has a finite value.  
A straightforward  example is when the average available energy is finite. In this case, the state of the environment is constrained to the subspace composed of energy eigenstates with less energy than this available amount.

With the assumption that the quantum echo dynamics in the environmental
degrees of freedom is associated with fully chaotic underlying classical dynamics, we extend 
the arguments in \cite{gorin2006} to obtain a relation between the time average of the decoherence functions and the 
\textit{effective dimension} of the environment's Hilbert space in the strong perturbation 
regime. Indeed, by expanding the initial state of the environment in the eigenbasis $\{\vert \psi_l\rangle\}$ of the unperturbed evolution operator $\hat U^\dagger _m$ or the eigenbasis $\{\vert\widetilde\psi\rangle\}$ of the perturbed 
evolution operator $\hat U^\dagger_n$ one obtains
\footnote{If the environment is a periodic time-dependent system these are the eigenstates of the
corresponding Floquet evolution operators.}:
\begin{eqnarray}
\hat\omega(0)&=&\sum_{l=1}^{N_{m}}\sum_{k=1}^{N_m}\omega_{lk}(0)| \psi_l\rangle\langle \psi_k|=
\label{expansao-omega-0}\\
&=&\sum_{l=1}^{N_{nm}}\sum_{k=1}^{N_{nm}}\widetilde\omega_{lk}(0)| \widetilde\psi_l\rangle\langle \widetilde\psi_k|,\label{expansao1-omega-0}
\end{eqnarray}
where $N_m$ and $N_{nm}$ represent the number of eigenstates of the evolution operators  
which significantly contribute to the sums, {\it i.e.,} $\omega_{lk}(0)\approx 0$ for
$l,k > N_m $ and $\widetilde\omega_{lk}(0)\approx 0$ for
$l,k > N_{nm} $  .  At any future time, the reduced density matrix of the environment $\hat\omega(t)$ can be spanned by $N_{\rm eff}$ linearly independent orthogonal eigenvectors
$\{| \psi_l\rangle\}$ or $\{| \widetilde\psi_l\rangle\}$, where
\begin{equation}
N_{\rm eff}\equiv\max_{nm}(N_m,N_{\rm nm}).\end{equation}
Therefore, $N_{\rm eff}$ can be interpreted as the  
effective dimension of the environment's Hilbert space for all times. In 
Appendix \ref{appendixA} we show that due to the classically chaotic underlying dynamics of the
environment, the strong coupling regime to the central system (and for $N_{\rm eff}\gg 1$), it is possible to relate the time average of the system decoherence functions to:
\beq
\label{mean-valueFmn}
\langle F_{nm}\rangle= \frac{C}{N_{\rm eff}} =
\tilde C
\hbar_{\rm{eff}}^{\gamma}\;\;,
\eeq
where $C$ is a constant that depends on $\hat \omega (0)$  (see Appendix \ref{appendixA}) 
and $\tilde C$ is related to $C$
once we know the specific relationship between the effective Planck constant $\hbar_{\rm{eff}}$ and
$N_{\rm eff}$ that can involve an exponent $\gamma$. 
For example, in the case of environments with an autonomous Hamiltonian $\hat H_e$ and
time-independent couplings to the central system, the dynamics associated with the unperturbed
Hamiltonians $\hat H_m$ conserve energy. In this case, following the well
known semiclassical prescription ({\it i.e., Weyl's rule} \cite{ozoriolivro}) one obtains  
$N_{\rm eff} \sim N_m \approx \nu_{E_m}/(2\pi\hbar_{\rm{eff}})^{\gamma}$, where
$ \nu_{E_m}$ is the volume inside the phase-space surface of constant energy
$E_m\sim {{\rm Tr}}[\hat H_m\hat \omega(0)]$, and $\gamma$ is the number of degrees of freedom of the environment.

The description of the long-time behavior of the decoherence functions $ F_{nm}(t)$ in the case of 
chaotic environments is completed with the analysis of the width of the fluctuation around the mean value.
In Appendix \ref{appendixA} we obtain,
\begin{align}
\Delta F_{nm} &\equiv \sqrt{\langle{F_{nm}^2}\rangle -\langle F_{nm}\rangle ^2}=&\nonumber\\
&= \frac{G}{N_{\rm eff}}=
\tilde G
\hbar_{\rm{eff}}^{\gamma}\;\;,&
\label{rmsd-Fmn}
\end{align}
where $G$ is a constant that depends on $\hat \omega (0)$  (see Appendix \ref{appendixA}) 
and $\tilde G$ is related to $G$
once we know the specific relationship between $\hbar_{\rm{eff}}$ and
$N_{\rm eff}$.
We point out that to arrive at relations \eqref{mean-valueFmn} and \eqref{rmsd-Fmn}, 
the approximation $N_{\rm eff} \gg1$ must be made. With these conditions fulfilled we can say that the above results only depend on the initial state of the environment $\hat \omega(0)$ through the 
values of $C$ and $G$.


\section{Decoherence and equilibration of a central system by a chaotic environment}
\label{sectionIV}

Here we show that the results of Sec. \ref{sectionIII} imply that
 in the semiclassical regime ($\hbar_{\rm{eff}}\ll 1$) a central system with a dephasing-type coupling to an environment with fully chaotic 
dynamics equilibrates to a 
state given by the time-averaged state of the central system, {\it i.e.,} 
\beq
\label{time-averaged-state}
\hat\rho_{\rm eq}=\langle \hat \rho(t)\rangle=
\sum_{n} A_{nn}  \ket n\bra n,
\eeq
where $ \hat \rho(t)$ is given in Eq.(\ref{rdm}) [see Appendix \ref{appendixB} for a derivation 
of Eq.(\ref{time-averaged-state})]. This occurs 
independently of the initial state of the environment and of the central system 
provided that the initial state
of the composite system is uncorrelated (as in Eq.(\ref{uncoupled-initial-state})). 

Generically, there is a back and forth of information flow between the central system 
and the environment as the reduced state of the system $\hat \rho(t) $
fluctuates around the totally decohered state
$\langle \hat \rho(t)\rangle$ in Eq. (\ref{time-averaged-state}). As a consequence of this non-Markovian character of the reduced dynamics of the central system, the degree of distinguishability between the evolved state  $\hat \rho(t) $ and $\langle \hat \rho(t)\rangle$ oscillates. 
In order to quantify this fluctuations 
we use the trace distance \cite{NielsenChuang}: 
\beq
D(\hat \rho(t),\hat\rho_{\rm eq})=
\frac{1}{2}{{\rm Tr}}_c\sqrt{[\hat \rho(t)-\hat\rho_{\rm eq}]^2}\;\;,
\eeq
which is a measure of the distinguishability between the two quantum states  \cite{Breuer}.
An upper bound to the trace distance is given by the Hilbert-Schmidt distance $D_{HS}$
\cite{hilbert-schmidt-dist},
\begin{align}
D(\hat \rho(t),\hat \rho_{eq})&\leq&
\frac{\sqrt{N_c}}{2}D_{HS}(\hat \rho(t), \hat \rho_{\rm eq})
\label{upper-bound0}\\
&=&\frac{\sqrt{N_c}}{2}\sqrt{\sum_{n\neq m}|A_{nm}|^2F_{nm}(t)} 
\label{upper-bound}
\end{align}
where $D_{HS}(\hat \rho(t),\hat \rho_{\rm eq})\equiv
\sqrt{{{\rm Tr}}_c[(\hat \rho(t)-\hat\rho_{\rm eq})^2]}$ and 
$N_c$ is the effective dimension of the Hilbert space of the \textit{central system}, which is given by 
the number of essentially non-zero terms in the expansion of the central system's initial state 
$\hat \rho(0)$ in the eigenbasis of $\hat H_c$. We calculate
the Hilbert-Schmidt distance in Eq.(\ref{upper-bound}) from the expressions for $\hat \rho(t)$ (\ref{rdm}), 
$\langle \hat \rho(t)\rangle$ (\ref{time-averaged-state}).
Using the concavity of the square-root function we can estimate an upper
bound for the time-averaged fluctuations:
\beq
\label{average-trace-distance}
\langle D(\rho(t),\hat\rho_{\rm eq}) \rangle \leq \hbar_{\rm{eff}}^{\gamma/2}
\frac{\sqrt{N_c \tilde C}}{2}\sqrt{\sum_{n\neq m}|A_{nm}|^2} \;\;,
\eeq
where we use the result in Eq.(\ref{mean-valueFmn}) for $\langle F_{nm}(t) \rangle$.
According to this result, the fluctuations, measured
by $\langle D(\hat \rho(t),\hat\rho_{\rm eq}) \rangle$, tend to zero in the semiclassical limit ($\hbar_{\rm eff}\rightarrow 0$).
This means that in this limit  the evolved state  $\hat \rho(t)$ becomes indistinguishable from the totally uncorrelated state in Eq.(\ref{time-averaged-state}).
In the semiclassical regime, the entanglement between the central system and the environment reaches its maximum possible value (which is determined by $\hat \rho(0)$). 

We can see this by noting that if the initial total composite state 
is $\hat \rho_{ce}(0)=\hat \rho(0) \otimes \hat \omega(0)$,
and for a dephasing-type coupling of the central system to the environment, the purity of the reduced evolved density matrix has 
the lower bound,
\begin{align}
{{\rm Tr}}_c[\hat \rho(t)^2]&=\sum_{n} |A_{nn}|^2&+ \sum_{n\neq m}
|A_ {nn}|^2 F_{nm}(t) \nonumber \\
&\geq \mathcal{R}_{\rm inv}(\hat \rho(0))&\;,
\label{lower-bound-purity} 
\end{align}
where $\mathcal{R}_{\rm inv}(\hat \rho(0))\equiv{{\rm Tr}}[\mbox{diag}\hat \rho(0)]=\sum_{n} |A_{nn}|^2
$ is the generalized inverse participation ratio \cite{gorin2006}  of the initial reduced density matrix
in the energy eigenbasis of the central
system.
The maximum entanglement between  the central system and the environment is attained when the lower bound  in Eq.(\ref{lower-bound-purity}) is reached. 
By inserting the result of Eq.(\ref{mean-valueFmn}) into Eq.(\ref{lower-bound-purity})
and taking the time-average, one obtains
\beq
\langle{{\rm Tr}}_c[\hat \rho(t)^2]\rangle
\stackrel{\hbar_{\rm{eff}}\rightarrow 0}{\longrightarrow }
\mathcal{R}_{\rm inv}(\hat \rho(0))\;\;,
\eeq
which means that maximum entanglement is expected in the semiclassical regime.


\section{Disentanglement by chaotic environments}
\label{sectionV}

We now consider two noninteracting central systems, $\hat H_{c_i}$ ($ i=1,2$), each coupled to a local environment $\hat H_{e_i}$ through a dephasing-type coupling,
\beq
\label{general-model-2}
\hat H = \hat H_{c_1}+ \hat H_{c_2}+ \hat H_{I_1}+ \hat H_{I_2}+ \hat H_{e_1}+\hat H_{e_2}\;\;,\eeq
where $\hat H_{I_i}=g_i\hat S_i\otimes \hat  V_i$, with 
$[ \hat H_{c_i}, \hat S_i]=0$. For simplicity we consider the 
two local environments to have  Hamiltonians, $\hat H_{e_1}$ and $\hat H_{e_2}$,  with the
same functional form. Thus, classically this corresponds to two different chaotic systems
with the same dynamics.
We stress here that the occurrence of local environments  is the most common  situation in proposals of quantum computation and quantum information processing. In this framework
the two central systems in Eq.(\ref{general-model-2})  can be considered as
noninteracting  qubits in the time interval separating the action of two consecutive conditional  logical gates
that involve an interaction between them.
Assuming that the two system Hamiltonians $\hat H_{c_1}$ and $\hat H_{c_2}$ have discrete and non-degenerate spectra,
we denote $\{\ket {n_i}\}$ the basis of common  eigenstates of $ \hat H_{c_i}$ and $\hat S_i$:
$\hat H_{c_i} \ket {n_i} = \varepsilon_{n_i} \ket {n_i}$ and
$\hat S_i \ket {n_i} = s_{n_i} \ket {n_i}$.
The total initial state is the tensor product:
$\hat \rho_{ce}(0)=\hat \omega(0) \otimes \hat \rho(0) \otimes \hat \omega(0)$
where $\hat \rho=\hat\rho_{c_1+c_2}$ is the entangled initial state of the bipartite central 
system and $\hat\omega(0)$ is the initial state of each local environment.

After tracing over the environmental degrees of freedom, one obtains the reduced density
matrix at time $t$ for the central bipartite system:
\bea
&\hat\rho(t)&={\rm Tr}_{e_1}{\rm Tr}_{e_2}\left[\hat \rho _{ce}(t)\right]\nonumber\\
&=&\sum_{n_1,m_1,n_2,m_2} A_{n_1m_1n_2m_2}  
 e^{\frac{-i}{\hbar}(\varepsilon_{n_1}-\varepsilon_{m_1}+\varepsilon_{n_2}-\varepsilon_{m_2})t}\nonumber\\
 &\times&
 f_{n_1m_1}(t)f_{n_2m_2}(t) \ket {n_1}\bra {m_1} \otimes \ket {n_2}\bra {m_2},
 \label{reduc-density-c1-c2-t}
\eea
where $f_{n_im_i}(t)={\rm Tr}_{e_i}\left[ \hat M_{n_im_i}(t) \hat \rho_{e_i}(0) \right]$
is the allegiance amplitude introduced in Eq.(\ref{fidelity1}), and 
$\hat M_{n_im_i}$ are the
echo operators in the Hilbert space of the local environments [see Eq.(\ref{echo-operator})].

Note that when $F_{n_im_i}\equiv |f_{n_im_i}|^2=0$, for all $n_i\neq m_i$, the 
initial entangled state $\hat\rho(0)$ becomes separable. But for $F_{n_im_i}\ne 0$ the two subsystems are not necessarily 
entangled. 
In order to illustrate how
the allegiance amplitudes $f_{n_im_i}$ can control the degree of entanglement 
of the central systems, we consider a simple case of two qubits with 
the free system Hamiltonians in Eq.(\ref{general-model-2}) given by
$\hat H_{c_i}=\hbar\omega\hat \sigma_{z_i}/2$ ($i=1,2$), where
$\hat \sigma_{z_i}$ are the Pauli $z$ operators whose eigenstates 
are $\{\ket 0, \ket1\}$ (``up'' and ``down'' states, respectively). 
 If we start with the entangled two-qubit state
 $\ket {\psi_c} =\left(\ket {0}\ket{+} + \ket {1}\ket{-}\right)/2$ (where $\ket{\pm}=\left(\ket{0}\pm\ket{1}\right)/\sqrt 2$), the evolved reduced density
 matrix in Eq.(\ref{reduc-density-c1-c2-t}) in the basis 
 $\{\ket {00}\ket{01}\ket{10}\ket{11}\}$ is, 
\bea
\frac{1}{4}\left(
\begin{array}{cccc}
 1 & f_{01}(t)e^{-i\omega t} & f_{01}(t)e^{-i\omega t}& - f^2_{01}(t) e^{-i2\omega t}\\
  & 1 &|f_{01}(t)|^2 & -f_{01}(t) e^{-i\omega t} \\
  & & 1 & -f_{01}(t) e^{-i\omega t}\\
  & h.c. &  & 1 
\end{array}
\right).
 \eea
A simple way to measure the entanglement of this state is to calculate its negativity \cite{vidal2002}, 
${\cal{N}}_E(\hat\rho)=\sum_j (|\lambda_j|-\lambda_j)/2$, where $\lambda_j$ are the eigenvalues of the partial transpose of $\hat\rho$. For $0\leq F_{01}(t)<1$, one obtains 
${\cal{N}}_E(\hat\rho)=\left[\left| \beta \right|-\left(\beta\right)\right]/8$
where $\beta\equiv -F_{01}(t)-2\sqrt{F_{01}(t)}+1$ . 
This means that when $F_{01}(t)$ decreases from $1$, the entanglement between the two qubits decreases until this function reaches a critical value $F_{01}^{(cr)}\equiv\sqrt{-1+\sqrt 2}$ (when  $N_E(\hat\rho)=0$) and from then on the two qubits are no longer entangled to each other. 
This is a simple example of the so called entanglement sudden death \cite{eberly2004} by a 
dephasing reservoir. 

If $F_{01}(t)$ were a monotonously decreasing function of time, we could be sure that from the time in which  $F_{01}(t)\leq F_{01}^{(cr)}$ 
the two qubits would be disentangled. But the disentanglement is also guaranteed even
if, after some time, $F_{01}(t)$ starts to fluctuate around  a mean value 
$\langle F_{01} \rangle < F_{01}^{(cr)}$  with the condition that the width $\Delta F_{01}(t)$
of this fluctuation is sufficiently small. As shown in Sec.(\ref{sectionIII}), this situation 
is  always satisfied  for chaotic  environments with few degrees of freedom where the mean
values  $\langle F_{n_im_i} \rangle$ and the magnitude of the fluctuations
 $\Delta F_{n_im_i}(t)$ go to zero in the limit of small effective Planck constant
 of the environment.
 
We can outline the disentanglement of the two central systems due to the action of the local chaotic environments if we assume that they are identical, {\it i.e.,} 
$\hat H_{c_1}$ and  $\hat H_{c_2}$ with identical functional form, 
and with a spectrum of eigenenergies with 
non-degenerate gaps \footnote{A spectrum $\{E_l\}$ has non-degenerate gaps if and only if the equation 
$E_f-E_j+E_k-E_l=0$ has as solutions only the cases $f=j$ and $k=l$ or $f=l$ and
$k=j$ or $f=j=k=l$ \cite{Linden2009}.
}. In this case it is straight forward to calculate the time-average of the evolved reduced state of the bipartite central system if the initial state has the form
$\hat \rho_{ce}(0)=\hat \omega(0) \otimes \hat \rho(0) \otimes \hat \omega(0)$
($\hat \rho=\hat\rho_{c_1+c_2}$ is the entangled initial state of the bipartite central 
system), which unlike the time-averaged reduced state 
in Eq(\ref{time-averaged-state})  is not
necessarily a separable state. Nevertheless, it can be written as:
\beq
\label{time-average-state-entangled}
\langle \hat \rho(t) \rangle =\langle \hat \rho(t) \rangle_d +
\hat O(\hbar_{\rm{eff}}^{\gamma}),
\eeq 
where 
\beq
\langle \hat \rho(t) \rangle_d\equiv \sum_{n_1}\sum_{n_2}A_{n_1n_1n_2n_2}
\ket {n_1}\bra {n_1} \otimes \ket {n_2}\bra {n_2}\;\;,
\eeq
is a disentangled state,  and 
\beq
\hat O \equiv \sum_{n_1}\sum_{n_2} 
A_{n_1n_2n_2n_1} \langle F_{n_1n_2}(t)\rangle 
\ket {n_1}\bra {n_2} \otimes \ket {n_2}\bra {n_1}\;\;.
\eeq
Note that using  Eq.(\ref{mean-valueFmn}) we can write
$\hat O(\hbar_{\rm{eff}}^{\gamma})\equiv
\hbar_{\rm{eff}}^{\gamma} \hat {\tilde O}$ where $Tr[\hat{\tilde O}]=0$. 
Following the same reason as in Section \ref{sectionIV} we can
estimate the time-averaged fluctuations of the evolved state $\hat \rho(t)$ 
around  $\langle \hat \rho(t) \rangle$ and we obtain:
$\langle D(\rho(t),\hat\rho_{\rm eq}) \rangle \leq  {\cal{O}}(\hbar_{\rm{eff}}^{\gamma})$.
Thus, we see that in the semiclassical regime $\langle \hat \rho(t) \rangle$ is the equilibrium 
state of the bipartite central system.
It is clear from Eq.(\ref{time-average-state-entangled}) that if the equilibrium state 
is not disentangled it is arbitrarily close to a separable state in the sense that a small
perturbation of order ${\cal{O}}(\hbar_{\rm{eff}})$ is enough to separate the state.
Indeed, if the perturbation is given  for example by the completely positive trace
preserving map: $\langle \hat \rho \rangle\rightarrow (1-\varsigma)\langle \hat \rho\rangle
+\varsigma(\langle \hat \rho\rangle-\hat {\tilde O})=\langle \hat\rho\rangle_d+\hat O-\varsigma\hat{\tilde O}$,  it is enough to apply a perturbation
of the order $\varsigma\approx \hbar_{\rm{eff}}^{\gamma}$ to obtain the disentangled state $\langle\hat\rho\rangle_d$.


\section{Numerical Simulations}
\label{sectionVI}

In order to confirm the analytical results in sections \ref{sectionIII} and \ref{sectionIV}, we perform
numerical simulations in which the environment is modeled by the periodically kicked Hamiltonian
\beq
\hat H_e= \frac{\hat P^2}{2M} + V_0 \cos(k_0\hat X)\sum_{n=0}^{\infty}(t/T-n)\;\;,
\eeq
where $[\hat X,\hat P]=i\hbar$. We consider two forms of dephasing-type coupling 
(see Eq.(\ref{general-model-1})) to a generic central system. The first is a linear coupling 
via the environment operator $\hat V=\hat P$ and the second is via the kicked coupling 
operator $\hat V=  \delta V_0 \cos(k_0\hat X)\sum_{n=0}^{\infty}(t/T-n)$.
Because the results we found for the two couplings are equivalent we only show those for the linear
coupling.

It is more convenient to work with the dimensionless coordinate $\theta=k_0\hat X$ 
and momentum $\hat p=k_0 T \hat P/M=\hbar_{\rm{eff}}\hat P/\hbar k_0$ that
satisfy the commutation relation $[\theta,\hat p]=i\hbar_{\rm{eff}}$ where
the effective Planck constant is $\hbar_{\rm{eff}}=k_0^2 \hbar T/M$.  Using these dimensionless variables and performing the energy transformation 
$
(\hbar_{\rm{eff}}T/\hbar)\hat H\rightarrow \hat H\;\;
$
to the total Hamiltonian $\hat H$ in Eq.(\ref{general-model-1}), the environment is represented by the kicked rotor Hamiltonian~\cite{Izrailev},
\beq
\hat H_{\rm KR}= \frac{\hat p^2}{2} + K \cos(\hat \theta)\sum_{n=0}^{\infty}(\tilde t-n)\;\;,
\eeq
where the new dimensionless kicking amplitude is $K=k_0^2T^2V_0/M=\hbar_{\rm{eff}}TV_0/\hbar$
and $\tilde t=t/T$. 

In the case of linear coupling $\hat V=\hat P$,  the echo-operator in Eq.(\ref{echo-operator}) 
is therefore associated with the dimensionless unperturbed Hamiltonian
\beq
\label{unperturbed-Hamil-p}
\hat H_m=\hat H_{\rm KR} + s_m \bar g \hat p\;\;,
\eeq
and the perturbed effective Hamiltonian 
\beq
\label{perturbed-Hamil-p}
H_n=\hat H_m + \epsilon_{nm}\hat p,
\eeq
where $\epsilon_{nm}\equiv \bar g (s_n-s_m)\equiv gk_0 T(s_m-s_n)$ is dimensionless.


Replacing the quantum operators $(\hat\theta,\hat p)$ with the classical coordinates
$(\theta,p)$ in $\hat H_{\rm KR}$, 
one obtains the classical Hamiltonian whose dynamics are governed by $K$. 
When $K$ increases from zero the phase space structure follows the Kolmogorov-Arnold-Moser (KAM) theory~\cite{lichter}
where the last invariant KAM torus is broken for $K=K_R\approx 0,97$, and the motion becomes unbound for $K>K_R$. If $K\sim 1$, the classical phase space is mixed, and for $K\gg 5$ the classical motion may be considered completely chaotic
having negligibly small stability islands~\cite{chirikov}.
The classical counterpart of the Hamiltonian $\hat H_m$ in 
Eq.(\ref{unperturbed-Hamil-p})  also presents essentially chaotic dynamics for $K\gg 5$  
even  when the strength of the linear 
coupling $s_m\bar g$ is large, because the linear coupling simply represents a linear shift in kicked system's classical phase space.  

We verified the results in Secs. \ref{sectionIII} and  \ref{sectionIV}  
in two different situations. The first is when the phase space of the kicked rotor is closed on the torus 
$-\pi\le\theta\le\pi$, $-\pi\le p\le\pi$ so that 
the Hilbert space of the environment 
is finite (dimension $N$). Because the angle variable $\theta$ is bounded, the 
quantum momentum eigenstates are discrete, $\hat p\ket j=\hbar_{\rm eff} j\ket j$ ($j=-N/2,\dots,0, N/2-1$), and $N$ is related to the effective Planck constant by 
$\hbar_{\rm{eff}}=2\pi/N$. When the underlying classical dynamics of the unperturbed 
Hamiltonians $\hat H_m$ in Eq.(\ref{unperturbed-Hamil-p}) are chaotic, at some finite relaxation time the environment's evolved state occupies all the 
available momentum eigenstates. It is thus clear that 
in this case the effective size of the environment's Hilbert space is $N_{\rm eff}=N=2\pi/\hbar_{\rm{eff}}$, for all $m$. Substituting this into Eqs.(\ref{mean-valueFmn}) and (\ref{rmsd-Fmn}) one obtains
\beq
1/N_{\rm eff}\propto \hbar_{\rm{eff}}
\Longrightarrow
\langle F_{nm}\rangle,\;\Delta F_{nm} \propto\hbar_{\rm{eff}}\;\;.
\label{Nm-hbar}
\eeq

\begin{figure}
\resizebox{\columnwidth}{!}{\includegraphics[width=80mm]{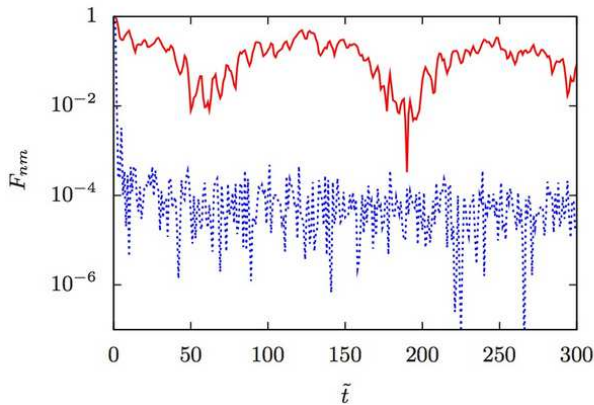}}
\caption{\label{fig3}
(Color online) Typical behavior of the decoherence function $F_{nm}(t)$ as a function of time when the environment has chaotic classical dynamics.  
In these plots the environment corresponds to the kicked rotor on a cylinder 
with $K=5$ and linear coupling to the central system. 
The unperturbed Hamiltonian in the corresponding echo operator is given in Eq.(\ref{unperturbed-Hamil-p}) with 
$s_m\bar g=0.1$ and the perturbed Hamiltonian is given by Eq.(\ref{perturbed-Hamil-p}) with the strong perturbation 
$\epsilon_{nm}=0.1$. The time is measured in number of kicks. 
 The full red line corresponds to $\hbar_{\rm{eff}}=0.92$ and the dashed blue line corresponds to $\hbar_{\rm{eff}}=0.015$.
The initial state of the environment  is $\hat \omega(0)=|p=0\rangle\langle p=0|$.}
\end{figure}

The second situation is the case in which the environment Hilbert space is infinite \footnote{We simulate this situation numerically by considering 
many more quantum levels than the size of the environment's effective Hilbert space so that the evolved state of the environment does not spread to the ``boundaries" of the defined momentum space.}.  
In this case the appearance  of an effective dimension of  the environment Hilbert space is well exemplified 
by the renowned phenomenon of dynamical localization\cite{Izrailev,Fishman1982}. It is well-known that  during the time interval $0<\tilde t<\tilde t_R$, the kicked rotor 
Hamiltonian $\hat H_K$  with $K\gtrsim 5$ and  $\hbar_{\rm{eff}}<<1$  presents diffusion in the discrete momentum levels in good correspondence with the classical model,   
{\it i.e.}, $ \Delta \hat p^2 (\tilde t) \approx D(K)\tilde t$ [where $D(K)$ is the classical diffusion coefficient]
\footnote{ Except for special values of $\hbar_{\rm{eff}}$ where the  energy growth 
is quadratic in time due to constructive quantum interference, characterizing the phenomenon called ``quantum resonance"  \cite{Izrailev}.}.
But after a finite relaxation time scale $\tilde t_R$, an important decrease in the diffusion
rate is observed until the state ceases to spread in momentum space. This happens
approximately  when the occupation number in momentum space of the evolved state 
reaches the value $N_{\rm eff}\equiv\max_{nm}(N_m,N_{\rm nm})$ determined by the initial state $\hat \omega(0)$ (see Sec. \ref{sectionIII}), {\it i.e.,}
\begin{equation}
\label{dyn-loc}
N_{\rm eff} \sim J \approx \sqrt{D(K)\tilde t_R}/\hbar_{\rm{eff}},
\end{equation} 
where
$J$, called localization length,  is essentially the width of the momentum eigenstates distribution.
Due to the  fully chaotic underlying classical dynamics, this phenomenon does not depend on 
the shape of the initial wave function (provided the initial state is not an eigenstate of the
evolution operator). 
When dynamical localization takes place, the mean level spacing between 
quasi-energy eigenstates involved in the expansion of the localized (in momentum)
initial wave function is $\Delta \approx 2\pi/J $. Thus, according to the Heisenberg principle, the minimum time
required for the dynamics to resolve this level spacing is $\tilde t_R\approx 1/\Delta \propto J$.
Substituting $\tilde t_R$  into  Eq.(\ref{dyn-loc}) one finally obtains
\beq
\label{Nm-hbar2}
1/N_{\rm eff} \sim 1/J \propto \hbar_{\rm{eff}}^2 
\Longrightarrow 
\langle F_{nm}\rangle ,\;
\Delta F_{nm} \propto\hbar_{\rm{eff}}^2\;\;.
\eeq
where we use the Eqs.(\ref{mean-valueFmn}) and (\ref{rmsd-Fmn}).

\begin{figure}
\begin{center}
\resizebox{\columnwidth}{!}{\includegraphics[width=80mm]{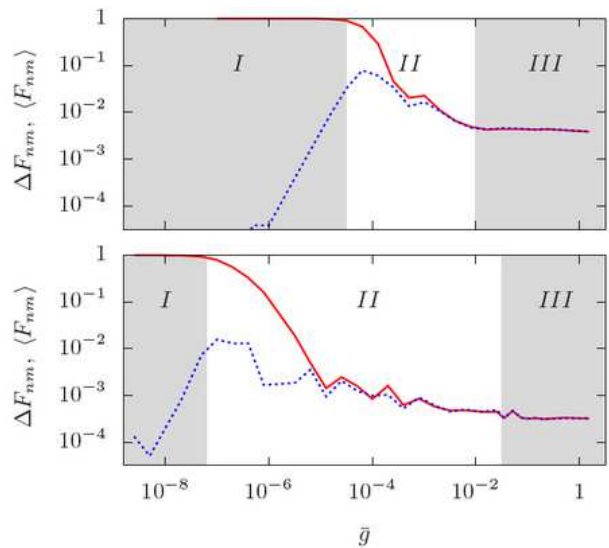}}
\end{center}
\caption{\label{fig4}
(Color online) Long-time behavior of the average $\langle F_{nm}(t)\rangle$ (red solid lines) and the 
standard deviation $\Delta F_{nm}$ (blue dashed lines), with $n=m+1$, as a function of the coupling strength to the environment. 
The environment was modeled by a kicked rotator
with $K=5$,  linearly  coupled to a central system with $s_m =1$.
The upper plot corresponds to the case where at large times the evolved state of the environment spreads over 
all the available Hilbert space (the kicked rotor is defined on a torus) and in the bottom plot we simulate the kicked rotor on a cylinder, where dynamical localization is observed.
Other parameters are $\hbar_{\rm{eff}}=0.025$ (top plot) and $\hbar_{\rm{eff}}=0.083$ (bottom plot) and the initial 
state of the environment is $\hat \omega(0)=|p=0\rangle\langle p=0|$.
The three regions identified in the plots are explained in the text.}
\end{figure}
The behavior in time of  the decoherence function $F_{nm}(t)$, determined by the echo dynamics 
with the conditional unperturbed Hamiltonian given in Eq.(\ref{unperturbed-Hamil-p}) with $K=5$, 
is illustrated in Figure \ref{fig3}, in which we consider a strong coupling between the central system 
and the environment.
In both the case in which at large times the evolved state of the environment spreads over
all the available Hilbert space and the case in which the available Hilbert space
is much greater than the dynamical localization length,
we found that at long-time scales, the decoherence function $F_{nm}(t)$ 
fluctuates around an asymptotic mean value $\langle F_{nm}(t)\rangle$. 
This is typical in the long-time behavior of 
the fidelity amplitude 
in systems with  fully classical chaotic dynamics in finite Hilbert spaces  \cite{gorin2006}.
We confirmed that this behavior is independent of the initial state of the environment, 
$\hat \omega(0)$.  

\begin{figure}
\resizebox{\columnwidth}{!}{\includegraphics[width=80mm]{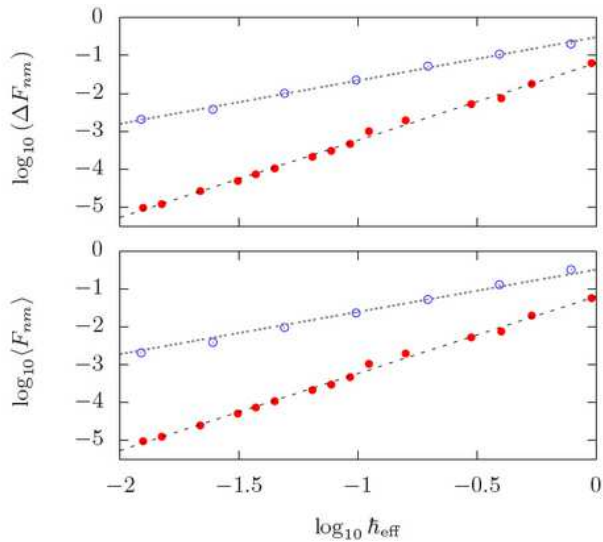}}
\caption{\label{fig5}
(Color online) $\langle F_{nm}\rangle $ (bottom graph) and $\Delta F_{nm}$ (top graph) as a function of the effective
Planck constant $\hbar_{\rm{eff}}$ for $n=m+1$.
The environment is a kicked rotator with $K=5$ with a strong linear coupling to the central system, 
$\epsilon_{nm}=0.1$. 
Blue circles correspond to a finite Hilbert space
and the red filled dots correspond to the case in which dynamical localization is observed. 
The linear fits are drawn. The corresponding slopes of these fits are in good agreement with analytical predictions and 
are given by: $2.03\pm0.03$ (top graph, filled red dots); 
$1.1\pm0.02$ (top graph, blue circles); 
$2.03\pm0.03$ (bottom graph, filled red dots); $1.08\pm0.02$ (top graph, blue circles). }
\end{figure}

The values of $\langle F_{nm}(t)\rangle$ and $\Delta F_{nm}$  for long times are plotted in Fig.\ref{fig4} as a function of coupling strength $\bar g$. The top graph corresponds to the situation in which the environment's phase space is closed on a torus and the bottom graph simulates the situation in which the environment has an infinite Hilbert Space and dynamical localization is observed.  
The red solid lines correspond to the numerical results for $\langle F_{nm}(t)\rangle$ and the blue dashed lines correspond to $\Delta F_{nm}$.
From these plots it is possible to identify three echo perturbation regimes: region {\bf I} is the weak perturbation regime and there is essentially no echo decay; Region {\bf II} is an intermediate regime in which the mean value and standard deviation of the decoherence function depend on both
$\hbar_{\rm{eff}}$ and the perturbation $\epsilon_{nm}=\bar g(s_n-s_m)$; and region {\bf III}  is the strong perturbation regime in which the mean value and standard deviation do not depend on the perturbation strength $\epsilon_{nm}$. It is in this third region
that the results in Eqs.(\ref{mean-valueFmn}) and (\ref{rmsd-Fmn}) apply
(with $C=G$ for the particular initial state $\hat \omega(0)$ considered).
In order to confirm these relations we plot in Fig.\ref{fig5} the numerical calculation of  $\log_{10}\langle F_{nm}\rangle$ (bottom graph) and $\log_{10}\left(\Delta F_{nm}\right)$ (top graph)
as a function of $\log_{10}\hbar_{\rm{eff}}$ for the kicked rotor closed on a torus (blue circles) and in the case in which the Hilbert space is much larger than the localization length $J$ (red filled dots). 
The linear fittings through the numerical points indeed confirm  the relations given in Eq.(\ref{Nm-hbar}) for the first case and the relations given in Eq.(\ref{Nm-hbar2}) for the second case. 
Therefore, in the long-time regime, when the environment has fully chaotic underlying classical dynamics and the system-environment coupling is strong, all off-diagonal matrix elements of the reduced density matrix in Eq.(\ref{rdm}) tend to zero in the semi-classical limit, \textit{i.e.,}
\begin{equation}
\langle F_{nm}\rangle,\;\Delta F_{nm}\stackrel{\hbar_{\rm{eff}}\rightarrow 0}{\longrightarrow }0.
\end{equation} 

\begin{figure}
\resizebox{\columnwidth}{!}{\includegraphics[width=80mm]{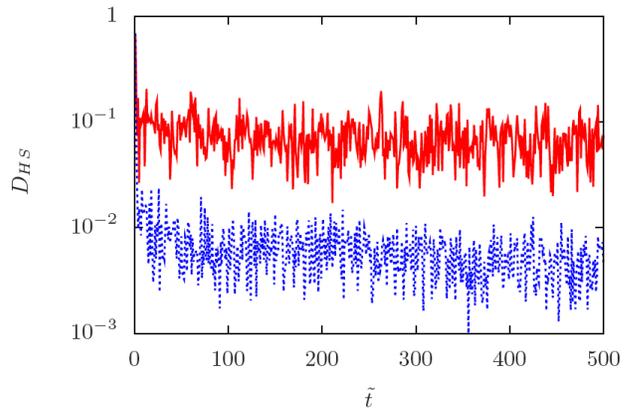}}
\caption{\label{fig5-1}
(Color online) Typical time evolution of the Hilbert-Schmidt distance $D_{HS}$ between the evolved 
reduced state of a central system $\hat \rho(t)$ and the totally decohered state 
$\langle\hat \rho\rangle$ in Eq.(\ref{time-averaged-state}) for $\hbar_{\rm eff}=0.53$ (red upper curve) and $\hbar_{\rm eff}=0.01$ (blue lower curve). In these graphs the central system is a harmonic oscillator and the system
plus environment (kicked rotor) initial state is $\hat\rho(0)\otimes \omega(0)$ where 
$\hat \rho(0)$ is a Schr\"oedinger-cat-like state and $\hat \omega(0)=|p=0\rangle\langle p=0|$.
}
\end{figure}

\begin{figure}
\resizebox{\columnwidth}{!}{\includegraphics[width=80mm]{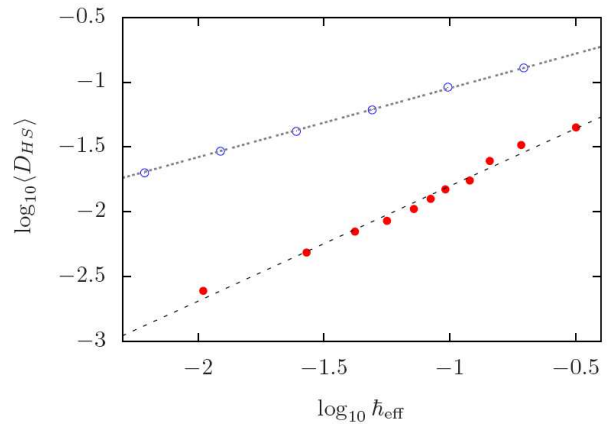}}
\caption{\label{fig6}
(Color online) Time-average value of the Hilbert-Schmidt distance $D_{HS}$ between the evolved 
reduced state of a central system $\hat \rho(t)$ and the totally decohered state 
$\langle\hat \rho(t)\rangle$ in Eq.(\ref{time-averaged-state}), as a function of
the effective Planck constant $\hbar_{\rm{eff}}$. 
The central system is a harmonic oscillator and the environment is a kicked rotator with $K=5$ on a torus (blue circles)
or on a ``cylinder" (red filled dots), in which case dynamical localization is observed. 
The lines correspond to linear fittings of the numerical results whose 
slopes are given by $0.9\pm0.1$ (fitting over filled red dots) and  $0.53\pm0.01$ (fitting over blue circles). 
The interaction between central system and environment is linear with $\epsilon_{nm}=0.1$ and the initial state of the system plus environment is the same as in Fig. \ref{fig5-1}. }
\end{figure}

In order to see the equilibration of the evolved reduced state of the central system $ \hat \rho(t)$
to the totally decohered state $\hat\rho_{\rm eq}=\langle \hat \rho(t)\rangle$ in Eq.(\ref{time-averaged-state}),
we consider as a central system a quantum harmonic oscillator
and an initial 
state $\hat \rho(0)=\ket \psi \bra \psi$ where 
$\ket \psi= {\cal N}\left(\ket \alpha + \ket {-\alpha}\right)$
($\ket \alpha $ a coherent state, {\it i.e}, an eigenstate of the
annihilation operator \cite{Cohen-Tannoudji}). We consider this Sch\"odinger-cat-like state initially uncorrelated from
the initial state of the environment $\hat \omega(0)=|p=0\rangle\langle p=0|$.
Equivalent results were obtained for different initial states $\hat \rho(0) \otimes \hat \omega(0)$.

The general behavior of the upper bound of the trace distance in Eq.(\ref{upper-bound0}) is shown in Fig.\ref{fig5-1}
where we see that the Hilbert-Schmidt distance $D_{HS}$ in long-time scales 
fluctuates around a mean value that decreases with $\hbar_{\rm eff}$.
In Fig.\ref{fig6}
we plot this mean value, given by  the time-average $\langle D_{HS}(\hat \rho(t),\hat\rho_{\rm eq})\rangle$, as a function of the effective Planck constant. 
According to the result in Eqs.(\ref{Nm-hbar}) and (\ref{Nm-hbar2}), the upper bound of the trace distance obtained in Eq.(\ref{average-trace-distance}) has a dependence on $\hbar_{\rm eff}^{1/2}$ 
when  the evolved state of the environment is spread over the entire available Hilbert Space 
and has a dependence on $\hbar_{\rm eff}$  when the evolved state dynamically localizes before filling the available Hilbert space. 
This is confirmed by the linear fit over the numerical points in Fig.(\ref{fig6}).
Hence, in the semiclassical limit we obtain for the trace distance:
\begin{equation}
\langle D(\hat \rho(t),\hat\rho_{\rm eq})\rangle\stackrel{\hbar_{\rm{eff}}\rightarrow 0}{\longrightarrow }0.
\end{equation}  
This means that in the semi-classical regime the central system always equilibrates to the totally
decohered state given in Eq. (\ref{time-averaged-state}), irrespective of the initial states of the system and the environment.


\section{conclusion}\label{sectionVII}
We studied the decoherence process of a generic quantum central system coupled via a dephasing-type interaction to an environment whose underlying classical dynamics is chaotic.
We have shown that if the environment has chaotic classical dynamics and the coupling to the central system is strong, the time-average of the decoherence functions and the width of the fluctuations around this time-average are inversely proportional to the effective Hilbert space dimension of the environment, defined as the dimension of the largest subspace onto which the projection of the initial state of the environment is not negligible, \textit{i.e.},  the maximum number of eigenstates of the perturbed or unperturbed conditional environment evolution operators needed to write the initial state of the environment. 
This means that if the effective Hilbert space of the environment is large, decoherence will occur, \textit{i.e.},
in the long-time regime the coherences of the reduced density matrix of the central system written in the preferential  basis will decrease significantly and on average will suffer only small fluctuations. If the quantum environment has chaotic underlying classical dynamics, the inverse of the its effective Hilbert space dimension will in general be proportional to some positive power of the effective Planck constant $\hbar_{\rm eff}$, which implies that decoherence of the central system is guaranteed in the semiclassical regime ($\hbar_{\rm eff}<< 1$). This is true even for an environment with few degrees of freedom. 
 We stress that, in general, if the coupling 
is not strong the central systems may have revivals of the coherences.

We thus confirm the intuitive notion that the fundamental quantity regarding the production of decoherence by chaotic environments is the dimension of the Hilbert space over which the environmental state is spanned, independent of the number of environmental degrees of freedom \cite{kohout}. In \cite{oliveira2009} the authors also discuss the connection between the effective Hilbert-space size and decoherence, but in our case it becomes clear that the decoherence is only guaranteed if the environment has chaotic underlying classical dynamics. 

For the system-environment models considered, in general the evolution of the reduced state of the system 
is not Markovian once the quantum information contained in the coherences of the reduced density matrix
in the preferential basis flows back and forth between system and environment. 
However, we showed that the decoherence process that occurs leads the central system
to equilibrate in the semiclassical limit to a diagonal state given by the time-average of its evolved reduced density matrix.
In this case, equilibration takes place independent of the initial state of the system and for a generic 
initial state of the environment (considered initially separable from the system).
This result is in agreement with that obtained in \cite{Linden2009}.
However, in our case, it is only the underlying chaotic classical dynamics 
of the environment that guarantees equilibration.

 We also  investigate the entanglement decay of two noninteracting central systems 
coupled via dephasing-type interactions to identical  quantum local  environments with classical chaotic dynamics.
In this case we show that the reduced evolved state of the system equilibrates to a non-diagonal 
state in the preferential basis, which in the semiclassical limit ($\hbar_{\rm eff} \rightarrow 0$)
is arbitrarily close to a diagonal state (in the preferential basis). 
Thus,  if the equilibrium state is not already disentangled, {\it i.e.}, the evolved reduced state presents entanglement sudden death, a very small perturbation is sufficient to lead the system into a separable state.


We confirmed all our analytical results with numerical simulations in which the environment is modeled by a kicked rotor in the regime where its underlying classical dynamics can be considered completely chaotic. 

\begin{acknowledgments}
We thank Adelcio Carlos de Oliveira, Arthur Rodrigo Bosco de Magalh\~aes, Rafael Chaves Souto Araujo and Raul Vallejos for helpful dicussions. We acknowledge financial support from the Brazilian funding agencies CAPES and CNPq.  This work was performed as part of the Brazilian project: ``Instituto Nacional de Ci\^{e}ncia e Tecnologia - Informa\c{c}\~{a}o Qu\^{a}ntica (INCT-IQ)''.
\end{acknowledgments}

\appendix
\section{Derivation of the results in Eqs.(\ref{mean-valueFmn}) and (\ref{rmsd-Fmn}) }
\label{appendixA}
In this appendix we calculate the time-average and standard deviation of the decoherence functions
$F_{nm}(t)$ assuming chaotic underlying classical dynamics of the environmental degrees of freedom. 
From here forward we assume $n\neq m$.  We expand the conditional unperturbed and perturbed evolution operators, $\hat U_m$ and $\hat U_n$ in their own eigenbasis:
\beq
\label{amplt-dec}
f_{nm}(t)={{\rm Tr}}\left[\hat U_m^{\dagger}\hat U_n \hat \omega(0) \right]
\eeq
\begin{eqnarray}
 \hat U_m^\dagger&=&\sum_{l=1}e^{\imath \xi_l t/\hbar}| \psi_l\rangle\langle \psi_l|,
 \label{expansao-u}\\
\hat{U}_n&=
&\sum_{l=1}e^{-\imath\widetilde \xi_l t/\hbar}|\widetilde \psi_l\rangle\langle\widetilde \psi_l|,
 \label{expansao-ut}
\end{eqnarray}
where $\xi_l$ and $\widetilde \xi_l$ are quasi-energies. To simplify notation, we do explicit the indexes $n$ and $m$ in 
$\xi_l $, $\widetilde \xi_l $, $|\psi_l\rangle$ and $|\widetilde\psi_l\rangle$.
The sums in Eqs.(\ref{expansao-u}) and (\ref{expansao-ut}) may go to infinity if the spectrums of evolution operators  are unbounded. 
We also expand 
the initial state of the environment in these eigenbases, as in Eqs.(\ref{expansao-omega-0})
and (\ref{expansao1-omega-0}). 
Inserting (\ref{expansao-u})  and (\ref{expansao-ut}) in (\ref{amplt-dec}) we obtain:
\beq
f_{nm}(t)=\sum_{l=1}\sum_{l'=1}e^{-i(\widetilde \xi_{l'}-\xi_l)t/\hbar}O_{ll'}B_{l'l}\;\;,
\eeq
where 
\bea
O_{ll'}&=&\langle \psi_l | \widetilde\psi_{l'} \rangle\\
\label{matrixAll}
B_{l'l}&=&
\sum_{k=1}^{N_m}\omega_{kl}(0)O^*_{kl'}=\sum_{k'=1}^{N_{nm}}\omega_{l'k'}(0) O^*_{k'l}\;\;.
\eea
Observing Eq.(\ref{matrixAll}) it is possible to see that matrix $B_{ll'}$ is essentially non-zero in the intervals $1\leq l\leq N_{m}$ and $1\leq l'\leq N_{nm}$. 
We remind the reader that $1\leq n,m\leq N_C$, where $N_C$ is the number of terms that significantly contribute to the expansion of the initial state of the central system $\rho(0)$ in the eigenbasis $\{|n\rangle\}$. Hence, as in Sec. \ref{sectionIII}, we use the definition of effective dimension of the environment 
subspace: 
$N_{\rm eff}\equiv\mbox{max}_{nm}(N_m,N_{nm})$, where the maximum is taken over all possible $n$ and $m$.
Indeed, at any time, the reduced density matrix of the environment can be written as 
\begin{eqnarray}
\hat\omega(t)&=&\sum_{l=1}^{{N_{\rm eff}}}\sum_{k=1}^{{N_{\rm eff}}}\omega_{lk}(0)
e^{-i(\xi_l-\xi_k)t/\hbar}| \psi_l\rangle\langle \psi_k|=
\\
&=&\sum_{l=1}^{{N_{\rm eff}}}\sum_{k=1}^{{N_{\rm eff}}}\widetilde\omega_{lk}(0)
e^{-i(\widetilde \xi_l-\widetilde \xi_k)t/\hbar}
|\widetilde\psi_l\rangle\langle \widetilde\psi_k|\;\;.
\end{eqnarray}
and the allegiance amplitude can be written as 
\beq\label{f(t)}
f_{nm}(t)=\sum_{l=1}^{N_{\rm eff}}\sum_{l'=1}^{N_{\rm eff}}e^{-i(\widetilde \xi_{l'}-\xi_l)t/\hbar}O_{ll'}B_{l'l}\;\;.
\eeq

For the time average of the decoherence functions we immediately obtain
\begin{eqnarray}
 \langle F\rangle&=&\sum_{l=1}^ {N_{\rm eff}}\sum_{l'=1}^{N_{\rm eff}}\vert B_{ll'}\vert^2\vert O_{ll'}\vert^2.\label{average}
\end{eqnarray}
We consider a non-degenerate chaotic quasi-energy spectra and strong enough perturbation such that the bases  $\{|\psi_l\rangle\}$ and $\{|\widetilde \psi_{l}\rangle\}$ are uncorrelated.  
In this case, 
one can assume the 
overlap matrix $\mathbb{O}$, formed by the elements $O_{ll'}$, to be a random unitary matrix~\cite{RMT}. 
In the limit of large $N_{\rm eff}$  (such that $(N_{\rm{eff}}+1)N_{\rm eff}\approx N_{\rm eff}^2$), the matrix elements $O_{ll'}$ can be considered complex random numbers with a Gaussian distribution  
$\propto\exp(-N_{\rm eff}|O_{ll'}|^2)$~\cite{CUE}. We can then average over this distribution of matrix elements:
\begin{eqnarray}
\overline{ O_{ll'}}&=&\overline{O_{ll'}^*}=0;\nonumber\\
\overline{\vert O_{ll'}\vert^2}&=&1/N_{\rm eff};\nonumber\\
\overline{\vert O_{ll'}\vert^4}&=&2/N_{\rm eff}^2,\label{averages}
\end{eqnarray}
where $\overline{\bullet}$ denotes the average over the Gaussian distribution of matrix elements. Using expressions (\ref{averages}) in 
(\ref{average}), for $N_{\rm eff}\gg1$ we finally obtain:
\begin{equation}
\overline{\langle F_{nm}\rangle}=\frac{C}{N_{\rm eff}},
\end{equation}
where $C \equiv \mathcal{R}_{\rm inv}(\hat \omega(0))+{\rm Tr}(\omega(0)^2)$ and $\mathcal{R}_{\rm inv}(\hat \omega(0))\equiv\sum_{l=1}^{N_{\rm eff}}\vert\omega_{ll}(0)\vert^2$ is the generalized inverse participation ratio \cite{gorin2006}  
of the initial reduced density matrix of the environment in the eigenbasis of the unperturbed 
evolution operator $\hat U_m$. 

In order to calculate $\Delta F_{nm}$ in Eq.(\ref{rmsd-Fmn}), we must calculate  $\langle F_{nm}^2\rangle$ 
starting from:
\begin{eqnarray}\label{variance2}
 F_{nm}^2&=&\sum_{f,j,k,l=1}^{N_{\rm eff}}\sum_{f',j',k',l'=1}^{N_{\rm eff}} 
e^{\frac{i}{\hbar}(\xi_f-\xi_{j}+\xi_k-\xi_{l}-\widetilde \xi_{f'}+\widetilde \xi_{j'}-\widetilde \xi_{k' }+\widetilde \xi_{l'})t}\nonumber\\
&\times& B_{ff'}^*B_{jj'}B_{kk'}^*B_{ll'} O_{ff'}O^*_{jj'}O_{kk'}O^*_{ll'}.
\end{eqnarray}
For chaotic quasi-energy spectra it is reasonable to assume that they 
have non-degenerate gaps \cite{gaps}
, so there exist only nine possible sets of conditions on $\{f,j,k,l\}$ and $\{f',j',k',l'\}$
for which the time average 
$ \langle e^{i(\xi_f-\xi_j+\xi_k-\xi_{l}-\widetilde \xi_{f'}+\widetilde \xi_{j'}-\widetilde \xi_{k' }+\widetilde \xi_{l'})t/\hbar}\rangle$  
is non-zero. In each case, one must compute the average over the Gaussian distribution of the elements of the overlap matrix $\mathbb{O}$ and
then sum the results. In the limit of large $N_{\rm eff}$, we obtain, 
\beq
\overline{\langle F_{nm}^ 2\rangle}=\overline{\langle F_{nm}\rangle}^2+\frac{2}{N_{\rm eff}^ 2}
\sum_{k=1}^{N_{\rm eff}}\sum_{l(\neq k)=1}^{N_{\rm eff}}|\omega_{ll}(0)|^2|\omega(0)_{kk}|^ 2+\frac{\varphi^ 2}{N_{\rm eff}^ 2}\;,
\eeq
and therefore Eq.(\ref{rmsd-Fmn}) where
$G \equiv\sqrt{2(\mathcal{R}_{\rm inv}(\hat\omega(0)))^ 2-2\sum_{l}|\omega_{ll}(0)|^4+\varphi^ 2},$ and 
$\varphi\equiv\sum_{l}\sum_{k} \left(|\omega_{ll}(0)|^2+|\omega_{kl}(0)|^2\right)$.

\section{The equilibrium state $\langle\rho(t)\rangle$}
\label{appendixB}
In this appendix we calculate the time average of the reduced system density matrix given in Eq.(\ref{rdm}).
The result is the state in Eq.(\ref{time-averaged-state}) that was shown that corresponds to the equilibrium state of the central system in the semiclassical regime.
We begin by expanding the conditional unperturbed operators $\hat U_m^\dagger$ and $\hat U_n$ as in (\ref{expansao-u}) and inserting these into Eq.(\ref{rdm}) to obtain
\beq
\hat\rho(t)=\sum_{n,m}\sum_l A_{nm}
e^{-\frac{i}{\hbar}(\Delta_{nml})t} 
\langle l |\hat\omega(0)| l\rangle |m\rangle\langle n|\;,
\eeq
where $\Delta_{nml}\equiv \varepsilon_n-\varepsilon_m+E_l^{(n)}-E_{l}^{(m)}$ and
$E_{l}^{(m)}$ are the quasi-energies of the unperturbed conditional evolution operator $\hat U_m$. 
Taking the time average defined in Eq.(\ref{time-averaged}) the only nonzero terms are those
where $\Delta_{nml}=0$, that for a non-degenerate spectra $\{\varepsilon_n\}$
only happens when $n=m$. So, we obtain
\begin{eqnarray}
\langle\hat\rho(t)\rangle&=&\sum_{n,m}A_{nm}\delta_{nm}\sum_l \langle l |\hat\omega(0)| l\rangle |m\rangle\langle n|\nonumber\\
&=&\sum_nA_{nn}|n\rangle\langle n|;
\end{eqnarray}
taking into account that ${\rm Tr}[\hat\rho(0)]=1$.


\end{document}